\documentclass[conference]{IEEEtran}
\IEEEoverridecommandlockouts
\usepackage{cite}
\usepackage{amsmath,amssymb,amsfonts}
\usepackage{algorithmic}
\usepackage{graphicx}
\usepackage{textcomp}
\usepackage{xcolor}

\usepackage{amsmath, amssymb, amsfonts, amsthm, booktabs, cancel, comment, enumitem, eurosym, float, graphicx, mathtools, multirow, nicefrac, placeins, soul, steinmetz, stfloats, siunitx, svg, url, xcolor, tabularx}%
\usepackage{commath}
\usepackage{cite}
\usepackage{subfigure}
\usepackage{hyperref}
\usepackage{nomencl}
\usepackage{cleveref}
\makenomenclature

\floatstyle{ruled}
\newfloat{model}{thp}{lop}
\floatname{model}{\textsc{Model}}

\makeatletter
\renewcommand*{\fnum@model}{\fname@model}
\makeatother

\def\BibTeX{{\rm B\kern-.05em{\sc i\kern-.025em b}\kern-.08em
    T\kern-.1667em\lower.7ex\hbox{E}\kern-.125emX}}
\begin{document}

% \title{Active Distribution Network Reconfiguration: Impacts on Aggregated DER Flexibility Provision
\title{Impacts of Distribution Network Reconfiguration on Aggregated DER Flexibility
% Conference Paper Title*\\
% {\footnotesize \textsuperscript{*}Note: Sub-titles are not captured in Xplore and should not be used}
\thanks{This work was carried out as a part of the ATTEST project (the Horizon 2020 research and innovation programme, grant agreement No 864298).}
}

% \and
% \and
% \and
% \and
% \and

\author{\IEEEauthorblockN{Andrey Churkin\textsuperscript{1},
Miguel Sanchez-Lopez\textsuperscript{1,2},
Mohammad~Iman~Alizadeh\textsuperscript{3},
Florin~Capitanescu\textsuperscript{3},\\
Eduardo~A.~Martínez~Ceseña\textsuperscript{1,4}, Pierluigi~Mancarella\textsuperscript{1,5}
\vspace{1\jot}}
\IEEEauthorblockA{\textsuperscript{1}Department of Electrical and Electronic Engineering, the University of Manchester, UK \\
\textsuperscript{2}Departamento de Ingenieria Electrica, Universidad de Chile, Chile\\
\textsuperscript{3}Luxembourg Institute of Science and Technology (LIST), Luxembourg\\
\textsuperscript{4}Tyndall Centre for Climate Change Research, UK\\
\textsuperscript{5}Department of Electrical and Electronic Engineering, the University of Melbourne, Australia\\
\{andrey.churkin; alex.martinezcesena; p.mancarella\}@manchester.ac.uk, \{mohammad.alizadeh; florin.capitanescu\}@list.lu}
}

\maketitle

\begin{abstract}
The ongoing integration of controllable distributed energy resources (DER) makes distribution networks capable of aggregating flexible power and providing flexibility services at both transmission and distribution levels. The aggregated flexibility of an active distribution network (ADN) can be represented as its feasible operating area in the P-Q space. The limits of this area are pivotal for arranging flexibility markets and coordinating transmission and distribution system operators (TSOs and DSOs). However, motivated by the current technical limitations of distribution networks (e.g., protection schemes), existing literature on ADN flexibility and TSO-DSO coordination mostly focuses on radial networks, overlooking the potential benefits of network reconfiguration. This paper, using a realistic meshed distribution system from the UK and the exact ACOPF model for flexibility estimation, demonstrates that network reconfiguration can increase the limits of ADN aggregated flexibility and improve the economic efficiency of flexibility markets.
\end{abstract}

\begin{IEEEkeywords}
Active distribution network (ADN), distributed energy resources (DER), flexibility services, network reconfiguration, TSO-DSO coordination.
\end{IEEEkeywords}

\section{Introduction}
Modern distribution networks incorporate increasing amounts of distributed energy resources (DER) and continuously improve their observability and controllability \cite{Eid2016}. With these controllable flexible resources, active distribution networks (ADNs) now offer attractive means to aggregate flexible power and provide flexibility services at both distribution and transmission levels. To enable coordination between transmission and distribution system operators (TSOs and DSOs) and arrange TSO-DSO flexibility markets, it is necessary to assess the limits of aggregated flexibility that ADNs can deliver \cite{Schittekatte2020,Givisiez2020,Sanjab2022}. This has motivated research on flexibility areas in the P-Q space to capture the aggregated ADN flexibility at the primary substation or TSO/DSO interface \cite{Silva2018,Contreras2018,Capitanescu2018,Riaz2021,Bolfek2021,Contreras2021}. However, despite the rapidly evolving research on ADN
flexibility, existing studies mostly focus on radial distribution networks, overlooking the potential benefits of network reconfiguration. There have been some attempts to quantify the value of ADN reconfiguration, e.g., in \cite{Capitanescu2014}. Yet, the impacts of network reconfiguration on aggregated DER flexibility and TSO-DSO coordination remain largely unexplored.

Distribution network reconfiguration is well known in power systems research as a tool for improving system security and optimising its operation, e.g., for loss reduction and voltage control \cite{Baran1989-2,NIKKHAH2023}. But, historically speaking, active network reconfiguration and meshed operation have not been economically attractive at the lowest voltage levels of the distribution networks, being typically considered only during emergency conditions \cite{SYRRI2016,MARTINEZCESENA2016}. As a result, existing assets (e.g., protections) are generally only suitable for radial operation, and existing literature on ADN flexibility has focused on radial networks \cite{Sanjab2022,Capitanescu2018,Riaz2021}. However, network reconfiguration, including meshed options, can become more attractive with increasing DER integration and the rising interest in reducing power losses and environmental concerns \cite{c2c}.

This paper demonstrates the benefits that network reconfiguration can bring in terms of increased ADN flexibility and improved economic efficiency of flexibility markets. For this purpose, a mixed integer quadratically constrained programming (MIQCP) model is developed to estimate the limits of flexibility P-Q areas and the cost of ADN aggregated flexibility under different network configurations. The impacts of the reconfiguration are illustrated using a meshed UK distribution network. The rest of the paper is structured as follows. Section~\ref{Section: Model} presents the proposed MIQCP flexibility estimation model. The case study, which highlights how ADN flexibility changes subject to different configurations, is presented in Section~\ref{Section: Results}. Finally, section~\ref{Section: Conclusion} concludes this work.

% Additional references from Miguel: \cite{en15145147}, \cite{9672091}, \cite{9910430}

\section{Modelling Framework}\label{Section: Model}
In this section, a mathematical programming model is proposed to analyse the effects of different network configurations on aggregated network flexibility (e.g., at the primary substation or TSO/DSO interface). The model presented in \eqref{Flex_OPF: objective1}-\eqref{Flex_OPF: constraints 2 q} is a single-stage (single-period) network flexibility estimation problem. This formulation is based on an exact AC optimal power flow model in rectangular voltage coordinates (ACROPF) \cite{Capitanescu2018}. Variables $e_k$ and $f_k$ are the real and imaginary rectangular voltage components at bus $k$ (i.e., $u_k = e_k + jf_k$), $p_{ij}$ and $q_{ij}$ stand for active and reactive power flows between buses $i$ and $j$, and $p_{k,g}$ and $q_{k,g}$ denote the active and reactive power of generators located in the network. The power of each flexible unit $f \in \mathcal{F}$ is given by its available P-Q upward and downward regulation capacities indicated by the corresponding arrows. It is worth noting that, in practice, the available upward and downward flexibility capacities would vary based on the initial operation of the flexible units before a flexibility service is requested. Finally, $x_{ij}$ are binary variables representing the status of lines (if lines are switched on or off).

%% initial formulation with flex. regulation binaries:
\begin{model}[t]
\caption{ACROPF for flexibility estimation \hfill [MIQCP]}
\label{Flex_OPF}
\begin{subequations} 
\vspace{-2\jot}
\begin{IEEEeqnarray}{llll}
% Variables:
\textbf{Variables:}\IEEEnonumber\\
e_k, f_k &\forall k \in \mathcal{K} \IEEEnonumber\\
p_{ij}, q_{ij}  &\forall (i,j) \in \mathcal{L} \IEEEnonumber\\
p_{k,g}, q_{k,g} &\forall k \in \mathcal{K}, \enskip \forall g \in \mathcal{G} \IEEEnonumber\\
\vspace{-3\jot}
p^\uparrow_{k,f},p^\downarrow_{k,f},q^\uparrow_{k,f},q^\downarrow_{k,f} &\forall k \in \mathcal{K}, \enskip \forall f \in \mathcal{F} \IEEEnonumber\\
\IEEEnonumber\\
\vspace{2\jot}
    x_{ij} \in \begin{dcases*}
        \{0,1\},  & \text{if reconfiguration line},\\
        1  &\text{other in-service lines}. 
        \end{dcases*} \IEEEnonumber\\
% Objective 1:
\textbf{Objective I:}\IEEEnonumber\\
\min \enskip {w_k^p}p_{k,g} + {w_k^q}q_{k,g} &k=k^{\text{ref}}
\vspace{2\jot}
\label{Flex_OPF: objective1} \IEEEyesnumber\\
% Objective 2:
\textbf{Objective II:}\IEEEnonumber\\
\min \enskip  \smashoperator{\sum_{k \in \mathcal{K}}} \smashoperator{\sum_{f \in \mathcal{F}}} C_{k,f}(p^\uparrow_{k,f},p^\downarrow_{k,f},q^\uparrow_{k,f},q^\downarrow_{k,f}) \vspace{2\jot} \label{Flex_OPF: objective2} &\IEEEyesnumber\\
% Constraints 1:
\textbf{Constraints I:}\IEEEnonumber\\
p_{ij} = ({e_i}^2+{f_i}^2)G_{ij} - (e_i e_j + f_i f_j)G_{ij} \label{Flex_OPF: p_ij}\\ \qquad\quad - (f_i e_j - e_i f_j)B_{ij} &\forall (i,j) \in \mathcal{L} \IEEEnonumber\\
q_{ij} = -({e_i}^2+{f_i}^2)B_{ij} + (e_i e_j + f_i f_j)B_{ij} \label{Flex_OPF: q_ij}\\ \qquad\quad - (f_i e_j - e_i f_j)G_{ij} &\forall (i,j) \in \mathcal{L} \IEEEnonumber\\
p_{k,g} - p_{k,d} + p^\uparrow_{k,f} - p^\downarrow_{k,f} \label{Flex_OPF: balance_p}\\
\qquad\quad-\smashoperator{\sum_{(k,j) \in \mathcal{L}}}p_{kj}x_{kj} = 0 &\forall k \in \mathcal{K}\IEEEnonumber\\
q_{k,g} - q_{k,d} + q^\uparrow_{k,f} - q^\downarrow_{k,f} \label{Flex_OPF: balance_q}\\
\qquad\quad-\smashoperator{\sum_{(k,j) \in \mathcal{L}}}q_{kj}x_{kj} = 0 &\forall k \in \mathcal{K}\IEEEnonumber\\
{p_{ij}}^2 + {q_{ij}}^2 \leq {S_{ij}^\text{max}}^2 &\forall (i,j) \in \mathcal{L} \label{Flex_OPF: Smax}\\
{v_k^\text{min}}^2 \leq {e_k}^2+{f_k}^2 \leq {v_k^\text{max}}^2 &\forall k \in \mathcal{K} \label{Flex_OPF: vmax}\\
{p_{k,g}}^\text{min} \leq p_{k,g} \leq {p_{k,g}}^\text{max} &\forall k, \enskip \forall g \label{Flex_OPF: p_lim}\\
{q_{k,g}}^\text{min} \leq q_{k,g} \leq {q_{k,g}}^\text{max} &\forall k, \enskip \forall g \label{Flex_OPF: q_lim}\\
0 \leq p^\uparrow_{k,f} \leq {p^{\uparrow\text{max}}_{k,f}} &\forall k, \enskip \forall f \label{Flex_OPF: p_up}\\
0 \leq p^\downarrow_{k,f} \leq {p^{\downarrow\text{max}}_{k,f}} &\forall k, \enskip \forall f \label{Flex_OPF: p_down}\\
0 \leq q^\uparrow_{k,f} \leq {q^{\uparrow\text{max}}_{k,f}} &\forall k, \enskip \forall f \label{Flex_OPF: q_up}\\
0 \leq q^\downarrow_{k,f} \leq {q^{\downarrow\text{max}}_{k,f}} &\forall k, \enskip \forall f \label{Flex_OPF: q_down}\vspace{2\jot}\\
% Constraints 2:
\textbf{Constraints II:}\IEEEnonumber\\
p_{k,g} = {p_{k,g}}^\prime &k=k^{\text{ref}} \label{Flex_OPF: constraints 2 p}\\
q_{k,g} = {q_{k,g}}^\prime &k=k^{\text{ref}} \label{Flex_OPF: constraints 2 q}
\end{IEEEeqnarray}
\end{subequations}
\end{model}

The model uses two sets of objective functions and constraints designed for two purposes: map the flexibility area limits or deploy the cheapest flexible units. Objective function \eqref{Flex_OPF: objective1} minimises or maximises network power consumption at a selected reference bus (e.g., primary substation or TSO/DSO interface). Coefficients $w_k^p$ and $w_k^q$ are introduced to control the optimisation directions and can be used to iteratively reconstruct the feasible P-Q flexibility area. Objective function \eqref{Flex_OPF: objective2} minimises the total cost of all flexible power regulations. It is used to find the optimal dispatch decisions and analyse the efficiency of the flexibility market for a given operating point. The first set of constraints, \eqref{Flex_OPF: p_ij}-\eqref{Flex_OPF: q_down}, defines the technical constraints of the network and flexible units. Active and reactive power flows are determined with \eqref{Flex_OPF: p_ij} and \eqref{Flex_OPF: q_ij}, where $G_{ij}$ and $B_{ij}$ are the conductance and susceptance of lines, respectively. Equations \eqref{Flex_OPF: balance_p} and \eqref{Flex_OPF: balance_q} represent active and reactive power balance for each node, where $p_{k,d}$ and $q_{k,d}$ are the loads at bus $k$. Line capacity limits and nodal voltage magnitude limits are imposed in \eqref{Flex_OPF: Smax} and \eqref{Flex_OPF: vmax}. The output of generators and flexible units is constrained in \eqref{Flex_OPF: p_lim}-\eqref{Flex_OPF: q_lim} and \eqref{Flex_OPF: p_up}-\eqref{Flex_OPF: q_down}. This set of constraints, \eqref{Flex_OPF: p_ij}-\eqref{Flex_OPF: q_down}, together with the objective function \eqref{Flex_OPF: objective1}, enables estimating the limits of the aggregated network flexibility and approximating it as the network feasibility boundary. When minimising the total flexibility cost with the objective function \eqref{Flex_OPF: objective2}, additional constraints \eqref{Flex_OPF: constraints 2 p}-\eqref{Flex_OPF: constraints 2 q} must be introduced to specify the selected feasible operating point.\footnote{Note that the cost-minimising model does not consider intertemporal constraints and switching costs. The inclusion of additional constraints and costs is the subject of future research.}

The presented ACROPF flexibility estimation formulation is a MIQCP problem. This problem is nonlinear and nonconvex, which imposes requirements on the solver: it must combine algorithms for both combinatorial optimisation (e.g., branch and bound) and nonlinear optimisation. Solving the MIQCP flexibility estimation problem can be intractable for an arbitrarily complex distribution network with an arbitrarily large number of flexible units. However, realistic networks can have a limited number of possible configurations. For example, as will be demonstrated in Section~\ref{Section: Results}, some distribution networks in the UK (typically 6.6 kV and 11 kV networks) have only one loop, a normally open point (NOP) connecting adjacent feeders. For such cases, the MIQCP flexibility estimation problem can be decomposed into continuous QCP subproblems corresponding to possible network configurations and solved with software for nonlinear continuous systems, such as Ipopt.

In the next section, the flexibility estimation model \eqref{Flex_OPF: objective1}-\eqref{Flex_OPF: constraints 2 q} will be applied to analyse the effects of network reconfiguration on the aggregated network flexibility. Changes in both the technical limits of network flexibility and the economics of flexibility provision will be traced and discussed. Additionally, the application of the exact ACROPF model to a realistic distribution network will demonstrate the nonlinearities and nonconvexities of ADN operation and highlight potential coordination issues of flexible units.

\section{Results and Discussion}\label{Section: Results}
\subsection{Case Study: 38-bus Synthetic Distribution Network}
\begin{figure}
    \centering
    \includegraphics[width=\columnwidth]{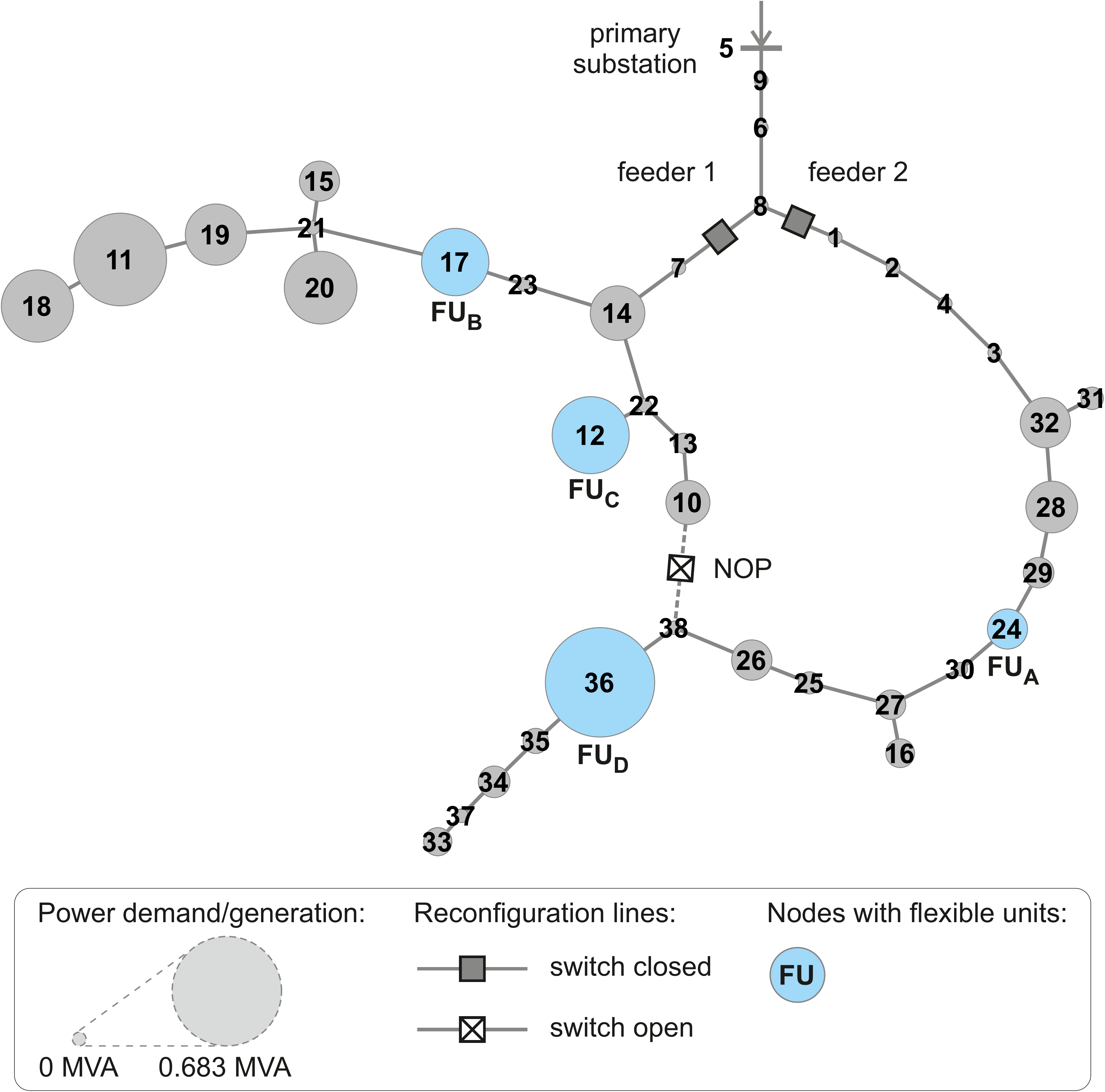}
    \caption{Case study: 38-bus distribution network with two feeders and four flexible units.}
    \label{Fig: Distribution_Network_UK}
\end{figure}

The impacts of network reconfiguration on the aggregated DER flexibility are illustrated with a 38-bus 6.6 kV UK distribution network \cite{network_UK}, which has been anonymised. The network is displayed in Fig.~\ref{Fig: Distribution_Network_UK} as a graph, where the sizes of the nodes represent power demand or generation at each bus, and lengths of the edges are proportional to the impedance between buses. Flexible operation of this network is challenging due to potential voltage problems and congestion. For example, feeder~2 has buses with high demand, which causes a significant voltage drop and restricts flexible power consumption. Conversely, feeder~1 has uncontrollable generation at buses 11, 17, 18, 19, and 20, which increases the voltage profile and dictates the power flows in the feeder. Four flexible units (denoted as FU in Fig.~\ref{Fig: Distribution_Network_UK}) are placed in the network. Their characteristics are listed in Table~\ref{table: data}. It is assumed that units can produce and consume power, which is consistent with DER technologies such as battery storage and flexible loads.

The network includes two radial feeders that can be interconnected through the NOP (typically for customer restoration purposes). That is, there are two sectionalising switches and one tie switch, which enables dynamic network reconfiguration. Excluding cases where some customers are isolated from the system, there are four possible network configurations to consider: (i) normal operation with NOP open (radial network), (ii) normal operation with NOP closed (meshed network), (iii) contingency power supply via feeder~1 (radial network, NOP closed, line 8-1 open), and (iv) contingency power supply via feeder~2 (radial network, NOP closed, line 8-7 open). The limited number of possible configurations makes it possible to decompose the flexibility estimation problem \eqref{Flex_OPF: objective1}-\eqref{Flex_OPF: constraints 2 q} into four subproblems, which will be explored in the following subsections. In this work, the ACROPF model was programmed in JuMP 1.4.0 for Julia 1.6.1 programming language, then decomposed into continuous QCP subproblems for the four configurations, and solved with Ipopt 3.14.4 solver.

\begin{table}[b]
\renewcommand{\arraystretch}{1.1}
\caption{Parameters of Flexible Units Placed in the Network}
\centering
\begin{tabular}{lllll}
\toprule %\hline
% \vspace{-1\jot}
\multirow{2}{*}{Parameter} & \multicolumn{4}{c}{Flexible unit}
\\
\cmidrule(l){2-5} 
% \vspace{-1\jot}
& A & B & C & D\\ 
\midrule %\hline
Bus \# & 24 & 17 & 12 & 36\\ 
P regulation limits (MW) & $\pm$1.0 & $\pm$1.0 & $\pm$1.0 & $\pm$1.0\\ 
Q regulation limits (MVAr) & $\pm$1.0 & $\pm$1.0 & $\pm$1.0 & $\pm$1.0\\
P cost (\$/MWh) & 375.0 & 350.0 & 325.0 & 300.0\\ 
Q cost (\$/MVArh) & 187.5 & 175.0 & 162.5 & 150.0\\
\bottomrule %\hline
\end{tabular}
\label{table: data}
\end{table}

\subsection{Aggregated Flexibility Limits}
\begin{figure}
    \centering
    \includegraphics[width=0.95\columnwidth]{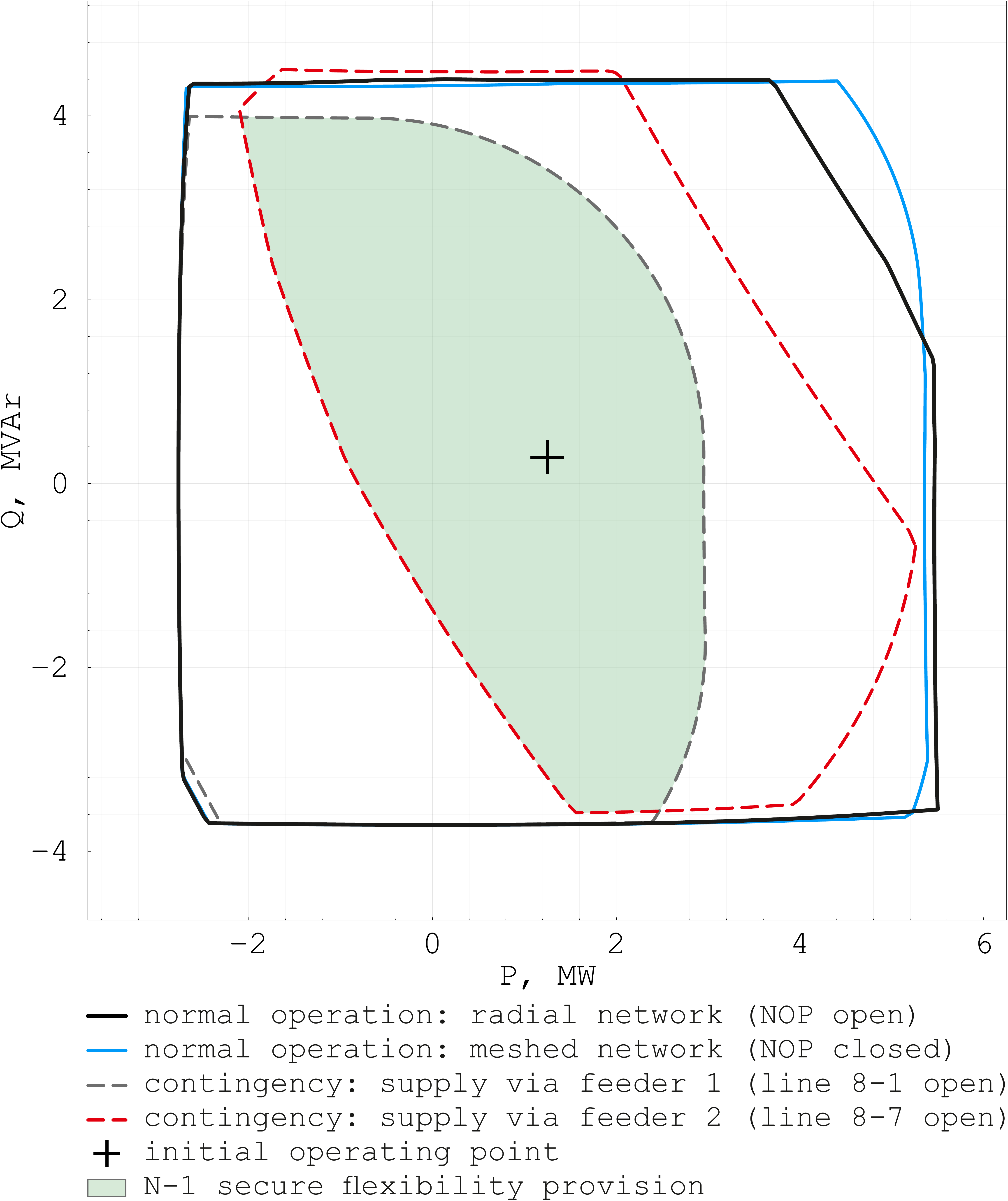}
    \caption{Boundaries of the feasibility areas for different network configurations.}
    \label{Fig: flex_boundaries}
\end{figure}

For each of the four considered network configurations, model \eqref{Flex_OPF: objective1}, \eqref{Flex_OPF: p_ij}-\eqref{Flex_OPF: q_down} was solved iteratively 200 times (for 200 extreme operating points, with a step of 0.08 MVA) to approximate the boundary of the network feasibility area at the primary substation. The flexibility areas estimated by model \eqref{Flex_OPF: objective1}, \eqref{Flex_OPF: p_ij}-\eqref{Flex_OPF: q_down} for the 38-bus system, subject to the four selected configurations, are shown in Fig.~\ref{Fig: flex_boundaries}. The coordinates represent the total active and reactive power consumption of the network, measured at the primary substation, and the cross marker corresponds to the initial operating point with all flexible units switched off. Note that the slightly curved horizontal and vertical lines correspond to unconstrained network operation where flexible units can fully deploy their P-Q regulation capabilities. Other inclined lines reflect the constrained network operation where flexible units cannot be fully activated. Considering normal network operation under radial and meshed topologies, the flexibility P-Q areas are similar for most operating points, except for the upper right part of the areas (increased active and reactive power consumption). When the NOP is open (radial case), the voltage profile of feeder~2 has a significant drop and is close to the lower voltage limit of 0.94 p.u. Therefore, in this case, flexible units A and D cannot fully increase their power consumption due to voltage constraints. When the NOP is closed (meshed case), voltages in feeder~2 increase which expands the limits of the aggregated flexibility. 
In addition to improving the voltage profile, meshing radial networks increases the capacity for transferring flexible power, making the operation of distant flexible units less constrained.
The difference in the feasibility areas demonstrates that network reconfiguration (meshing in this case), besides offering options to manage power losses and voltages, can be used to maximise aggregated flexibility.

The results show that the flexibility area greatly decreases in the last two cases (contingencies when the network is supplied only with feeder 1 or 2). This happens due to voltage problems at busses 18 and 33 and congestion of lines 7-14 and 3-32. The ``N-1 secure" flexibility area can be estimated as the intersection of areas under different network configurations, as shown in Fig.~\ref{Fig: flex_boundaries} with a green polygon. Considering the effects of network reconfiguration, DSO can impose restrictions on some flexibility services to guarantee the security of flexible power provision and ADN operation against contingencies.

\subsection{Economics of Flexibility Provision}
Different network configurations can affect the economics of flexible power provision. To explore these economic effects, model \eqref{Flex_OPF: objective2}-\eqref{Flex_OPF: constraints 2 q} was solved for 25,000 feasible operating points (with a step of 0.05 MVA). The simulations were performed for both the radial and meshed cases. Only normal network operation was considered in the economic analysis of flexibility provision, while configurations corresponding to contingencies were omitted.

\begin{figure*}
    \centering
    \includegraphics[width=0.95\textwidth]{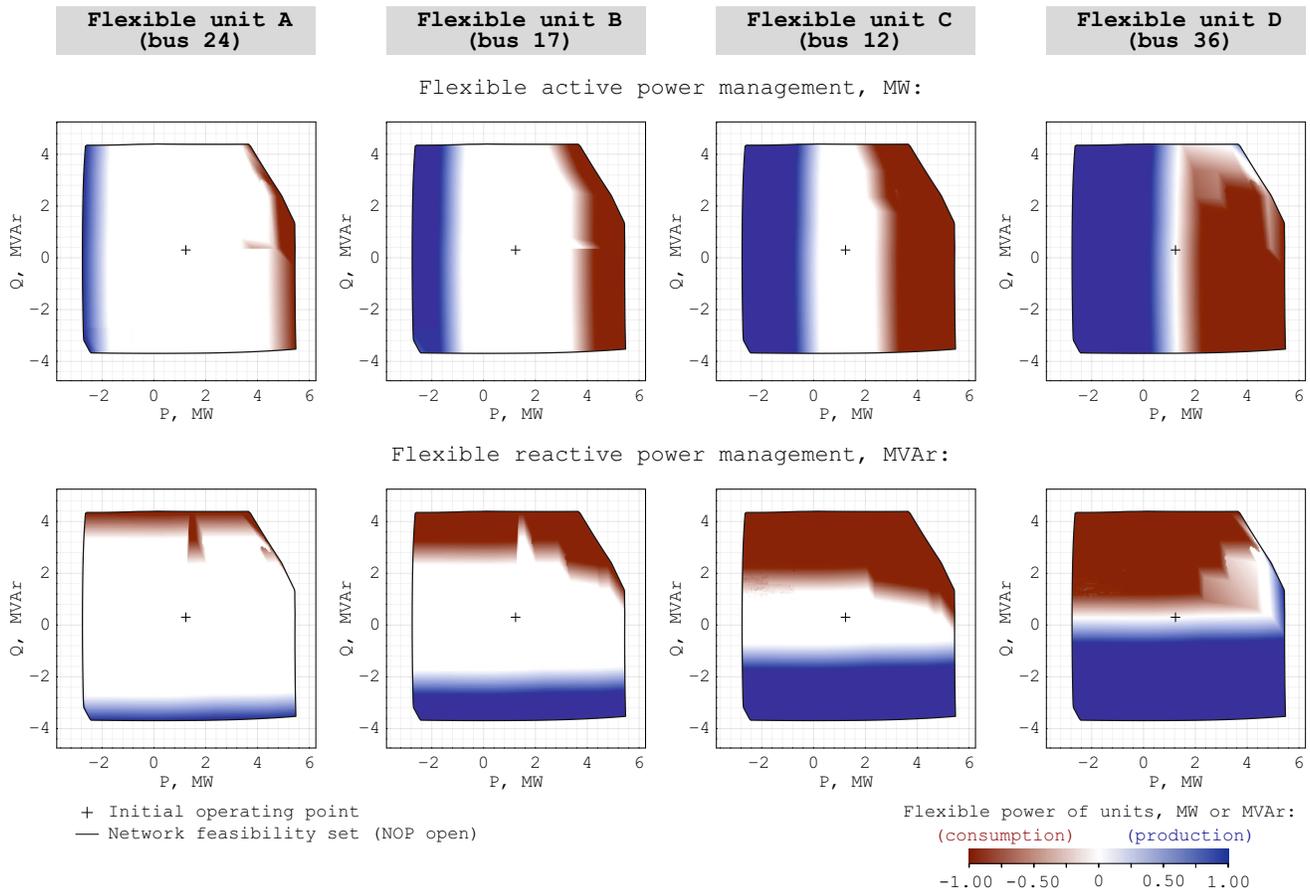}
    \caption{P-Q maps of the optimal flexible power regulations for the radial network topology (NOP open). The red-blue color scheme indicates operating points where flexible units consume or produce active and reactive power, in MW and MVAr.
    }
    \label{Fig: PQ_analysis_NOP_open}
\end{figure*}

The optimal (least-cost) operation of the flexible units under the radial network topology (NOP open) is visualised in Fig.~\ref{Fig: PQ_analysis_NOP_open} as P-Q maps of flexible power regulations. For each operating point, the maps specify which units should produce or consume flexible power to minimise the total cost of flexibility service, subject to the relevant voltage and thermal network constraints. For example, as expected, the cheapest unit, unit D, is activated for most flexibility service requests (e.g., a large portion of the flexibility area is highlighted in dark red and blue colours). However, this is not the case when the aggregated flexibility requests correspond to high power consumption (the top right areas highlighted in white or light red colours). At such points, the voltage level at bus 33 drops down to 0.94 p.u., which makes further flexible power consumption of unit D infeasible. Under such stressed conditions, other (more expensive) units have to consume flexible power, while unit D decreases its consumption or even starts producing power to alleviate voltage problems. This complex behavior is displayed in Fig.~\ref{Fig: PQ_analysis_NOP_open} as nonlinear changes in the power output of the units.\footnote{The observed complex nonlinear behavior of the flexible units is not the product of numerical instability or solver convergence. The results have been verified by other OPF formulations and solvers, e.g., for the radial network configuration, the same results were obtained with the DistFlow OPF model solved with Gurobi 10.0.} 

The nonlinear, sometimes abrupt, changes in the optimal power output of the flexible units pose two key challenges to the cost-effective operation of ADN and flexibility markets. Firstly, in practice, some units have ramp constraints that would prevent deploying the optimal portfolio of units, and would encourage the use of security margins to prevent infeasible operation (e.g., exceeding voltage or thermal limits). Secondly, information exchanges and strong coordination between units would be required to facilitate the complex unit coordination required to provide some flexibility services, especially under stressed network conditions (e.g., in the example above, when unit D is used to manage voltage constraints). Without such coordination, for instance, in decentralised peer-to-peer flexibility markets, DSOs would provide much more conservative flexible P-Q support for transmission systems.

\begin{figure*}
    \centering
    \includegraphics[width=0.95\textwidth]{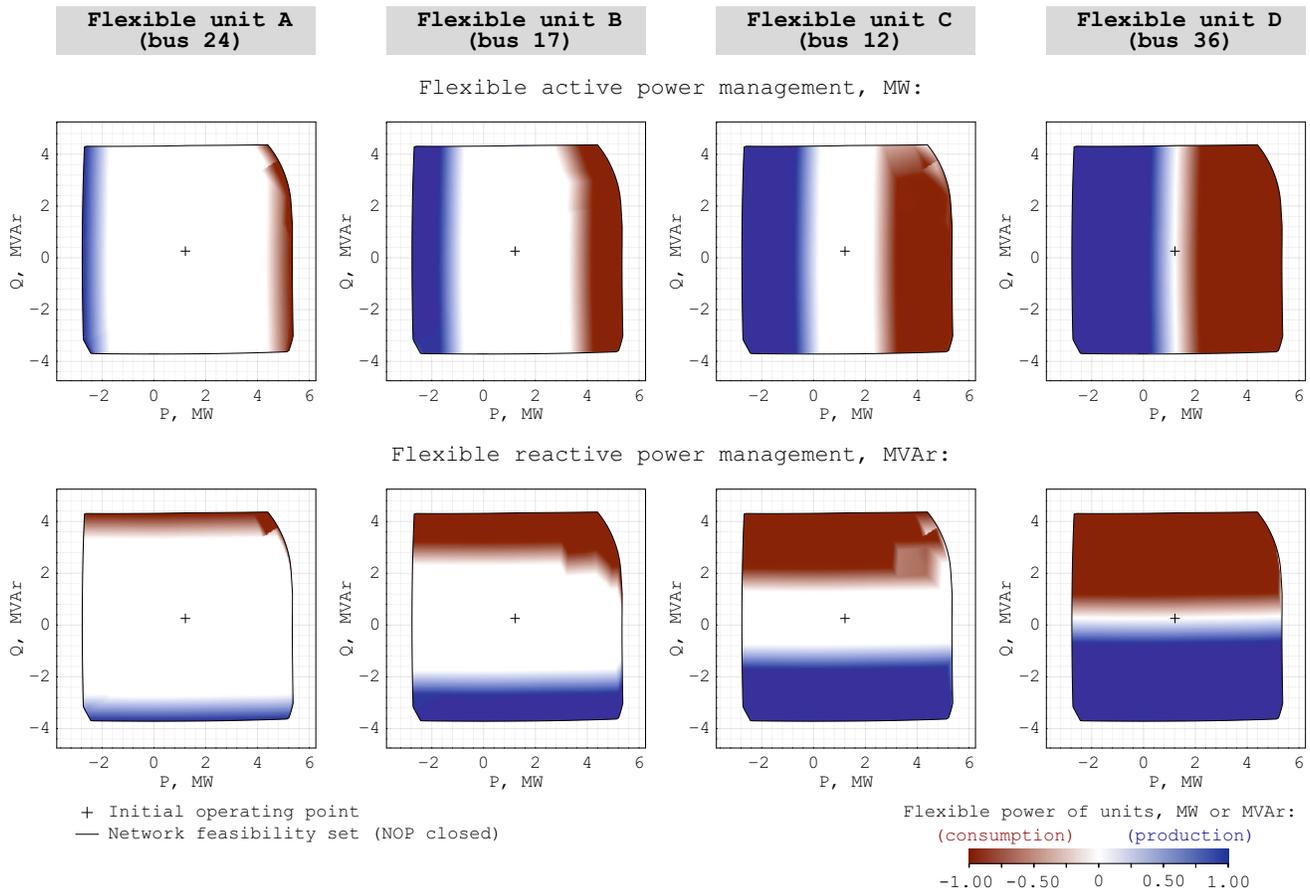}
    \caption{P-Q maps of the optimal flexible power regulations for the meshed network topology (NOP closed). The red-blue color scheme indicates operating points where flexible units consume or produce active and reactive power, in MW and MVAr.
    }
    \label{Fig: PQ_analysis_NOP_closed}
\end{figure*}

\begin{figure}
    \centering
    \includegraphics[width=0.95\columnwidth]{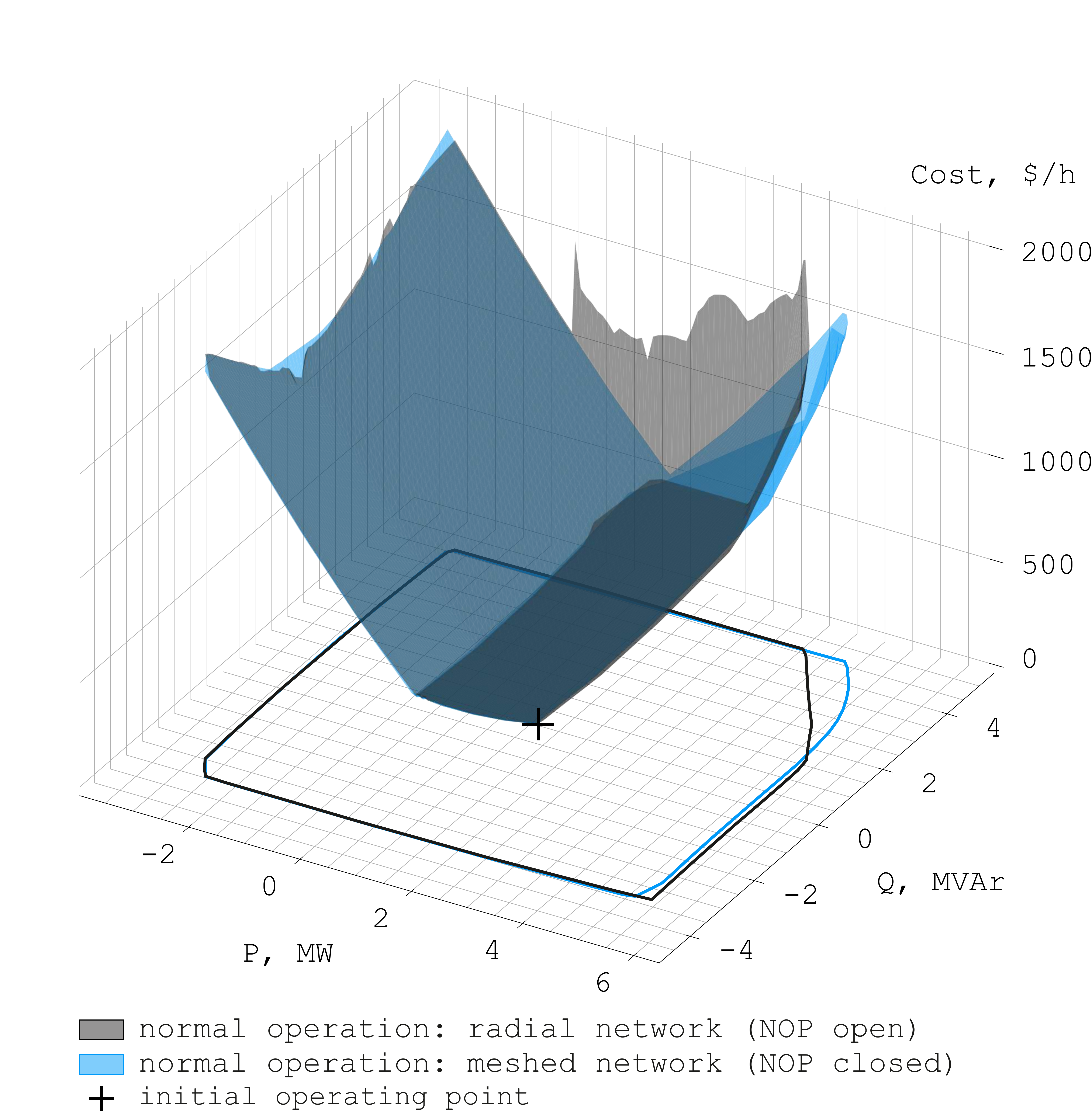}
    \caption{Total flexibility costs for different network configurations, \$/h.}
    \label{Fig: cost_surfaces}
\end{figure}

The P-Q maps of the optimal flexible power regulations for the meshed network topology (NOP closed) are presented in Fig.~\ref{Fig: PQ_analysis_NOP_closed}. Compared to the results for the radial network configuration, the cheapest unit D can now fully perform flexible power regulation for most feasible operating points (e.g., a larger portion of the area, compared with the radial case, is highlighted in dark red and blue colours). The voltage problems constraining the operation of unit D are now alleviated by network reconfiguration, making flexibility services cheaper and the flexibility market more efficient. The nonlinear rapid changes in the unit dispatch observed in Fig.~\ref{Fig: PQ_analysis_NOP_open} are now smoothed out and shifted to more expensive units. It follows that network reconfiguration can be used to reduce the cost of flexibility services and ease network operation and coordination between flexible units. A comprehensive cost analysis of aggregated flexibility under different network configurations is given in Fig.~\ref{Fig: cost_surfaces}, where costs for feasible operating points are displayed as surfaces. The flexibility cost under the radial network operation (grey surface) increases rapidly and unevenly for operating points with high power consumption. Conversely, the cost function corresponding to the meshed network operation (blue surface) grows slower and does not have cost spikes. The difference between the two surfaces illustrates the potential improvement in the flexibility market efficiency due to network reconfiguration.

\subsection{Discussion: Computational Aspects and Applicability}
In the above simulations, the computational challenges of the aggregated flexibility estimation problem were overcome by decomposing it into a small number of subproblems and fixing corresponding binary reconfiguration variables. However, the proposed ACROPF model \eqref{Flex_OPF: objective1}-\eqref{Flex_OPF: constraints 2 q} is generally hard to solve for meshed ADNs with multiple binary reconfiguration variables and numerous flexible units \cite{Capitanescu2017}. Such MIQCP models pose several challenges for solvers, such as difficulties in finding a feasible solution (an incumbent solution), convergence issues, and the inability to guarantee the global optimum. For example, for the considered 38-bus weakly meshed network, Gurobi 10.0 is able to solve the cost-minimising problem correctly for some operating points, but cannot correctly estimate the limits of the aggregated flexibility (the feasibility area boundary). Larger systems with more reconfiguration variables can create intractable problems. Therefore, there is a need for testing different optimisation algorithms, ACOPF formulations and their approximations for meshed ADNs with flexible units. A discussion of flexibility models programming and numerical issues can be found in \cite{Bolfek2021,Contreras2021,Capitanescu2014,Capitanescu2017,Lopez2021}. Note that existing ACOPF approximations and relaxations should be used with caution, as they can lead to inaccurate solutions and overestimation of ADN aggregated flexibility. Future research can focus on developing a tractable and accurate tool for estimating aggregated flexibility in ADNs, considering the effects of network reconfiguration. Additionally, the effects of including intertemporal constraints and switching costs should be explored.

Deploying the proposed flexibility estimation framework in real distribution systems and running the flexibility market will require overcoming several practical challenges. First, the security of flexibility provision and ADN operation must be guaranteed. DSO cannot commit to flexibility services that put a distribution network in a highly-stressed unreliable operation. Reliability analysis must be performed to estimate the impacts of potential contingencies or flexible units not delivering the right amount of flexible power. Second, it is necessary to introduce an information exchange system to coordinate the actions of flexible units. As demonstrated by the simulations, some units might need to perform fast regulations or produce flexible power to alleviate network constraints. Lack of unit coordination can result in an infeasible network operation and power outage. Finally, relay protection schemes must be upgraded to enable the operation of multiple flexible units in meshed ADNs.

\section{Conclusion}\label{Section: Conclusion}
This paper explores the operation of ADNs with flexible resources and demonstrates the effects of different network configurations on the aggregated DER flexibility. An accurate ACROPF flexibility estimation model, formulated as a MIQCP problem, is applied to a weakly meshed distribution network with two feeders in the UK. Extensive numerical simulations are performed for different network configurations to compare the limits of the aggregated flexibility and the economic efficiency of flexibility provision. The results illustrate that changing network topology (in this case from radial to meshed) enables increasing aggregated DER flexibility and reducing the cost of flexible power. Therefore, DSOs can use network reconfiguration as a tool to actively manage and optimise available flexible resources. However, to deploy this approach, multiple challenges have to be solved in the future, such as the security of flexible power provision under different network configurations, information exchange and coordination between relay protection and flexible units, and computational issues of flexibility estimation models.

% \bibitem{b1} G. Eason, B. Noble, and I. N. Sneddon, ``On certain integrals of Lipschitz-Hankel type involving products of Bessel functions,'' Phil. Trans. Roy. Soc. London, vol. A247, pp. 529--551, April 1955.
% \bibitem{b2} J. Clerk Maxwell, A Treatise on Electricity and Magnetism, 3rd ed., vol. 2. Oxford: Clarendon, 1892, pp.68--73.
% \bibitem{b3} I. S. Jacobs and C. P. Bean, ``Fine particles, thin films and exchange anisotropy,'' in Magnetism, vol. III, G. T. Rado and H. Suhl, Eds. New York: Academic, 1963, pp. 271--350.
% \bibitem{b4} K. Elissa, ``Title of paper if known,'' unpublished.
% \bibitem{b5} R. Nicole, ``Title of paper with only first word capitalized,'' J. Name Stand. Abbrev., in press.
% \bibitem{b6} Y. Yorozu, M. Hirano, K. Oka, and Y. Tagawa, ``Electron spectroscopy studies on magneto-optical media and plastic substrate interface,'' IEEE Transl. J. Magn. Japan, vol. 2, pp. 740--741, August 1987 [Digests 9th Annual Conf. Magnetics Japan, p. 301, 1982].
% \bibitem{b7} M. Young, The Technical Writer's Handbook. Mill Valley, CA: University Science, 1989.
% \end{thebibliography}
% \vspace{12pt}

\bibliographystyle{IEEEtran}
\bibliography{references.bib}

\begin{thebibliography}{10}
\providecommand{\url}[1]{#1}
\csname url@samestyle\endcsname
\providecommand{\newblock}{\relax}
\providecommand{\bibinfo}[2]{#2}
\providecommand{\BIBentrySTDinterwordspacing}{\spaceskip=0pt\relax}
\providecommand{\BIBentryALTinterwordstretchfactor}{4}
\providecommand{\BIBentryALTinterwordspacing}{\spaceskip=\fontdimen2\font plus
\BIBentryALTinterwordstretchfactor\fontdimen3\font minus
  \fontdimen4\font\relax}
\providecommand{\BIBforeignlanguage}[2]{{%
\expandafter\ifx\csname l@#1\endcsname\relax
\typeout{** WARNING: IEEEtran.bst: No hyphenation pattern has been}%
\typeout{** loaded for the language `#1'. Using the pattern for}%
\typeout{** the default language instead.}%
\else
\language=\csname l@#1\endcsname
\fi
#2}}
\providecommand{\BIBdecl}{\relax}
\BIBdecl

\bibitem{Eid2016}
C.~Eid, P.~Codani, Y.~Perez, J.~Reneses, and R.~Hakvoort, ``{Managing electric
  flexibility from Distributed Energy Resources: A review of incentives for
  market design},'' \emph{Renewable Sustain. Energy Rev.}, vol.~64, 2016.

\bibitem{Schittekatte2020}
T.~Schittekatte and L.~Meeus, ``{Flexibility markets: Q\&A with project
  pioneers},'' \emph{Utilities Policy}, vol.~63, 2020.

\bibitem{Givisiez2020}
A.~G. Givisiez, K.~Petrou, and L.~F. Ochoa, ``{A Review on TSO-DSO Coordination
  Models and Solution Techniques},'' \emph{Electr. Power Syst. Res.}, vol. 189,
  2020.

\bibitem{Sanjab2022}
A.~Sanjab, H.~Le~Cadre, and Y.~Mou, ``{TSO-DSOs Stable Cost Allocation for the
  Joint Procurement of Flexibility: A Cooperative Game Approach},'' \emph{IEEE
  Transactions on Smart Grid}, vol.~13, no.~6, pp. 4449--4464, 2022.

\bibitem{Silva2018}
J.~Silva, J.~Sumaili, R.~J. Bessa, L.~Seca, M.~A. Matos, V.~Miranda,
  M.~Caujolle, B.~Goncer, and M.~Sebastian-Viana, ``{Estimating the Active and
  Reactive Power Flexibility Area at the TSO-DSO Interface},'' \emph{IEEE
  Trans. Power Syst.}, vol.~33, no.~5, 2018.

\bibitem{Contreras2018}
D.~A. Contreras and K.~Rudion, ``{Improved assessment of the flexibility range
  of distribution grids using linear optimization},'' in \emph{Proc. 20th Power
  Systems Computation Conference, PSCC}, 2018.

\bibitem{Capitanescu2018}
F.~Capitanescu, ``{TSO–DSO interaction: Active distribution network power
  chart for TSO ancillary services provision},'' \emph{Electr. Power Syst.
  Res.}, vol. 163, 2018.

\bibitem{Riaz2021}
S.~Riaz and P.~Mancarella, ``Modelling and characterisation of flexibility from
  distributed energy resources,'' \emph{IEEE Trans. Power Syst.}, vol.~37,
  no.~1, pp. 38--50, 2022.

\bibitem{Bolfek2021}
M.~Bolfek and T.~Capuder, ``{An analysis of optimal power flow based
  formulations regarding DSO-TSO flexibility provision},'' \emph{Int. J.
  Electr. Power Syst.}, vol. 131, 2021.

\bibitem{Contreras2021}
D.~A. Contreras and K.~Rudion, ``{Computing the feasible operating region of
  active distribution networks: Comparison and validation of random sampling
  and optimal power flow based methods},'' \emph{IET Generation, Transmission
  and Distribution}, vol.~15, no.~10, 2021.

\bibitem{Capitanescu2014}
F.~Capitanescu, L.~F. Ochoa, H.~Margossian, and N.~D. Hatziargyriou,
  ``Assessing the potential of network reconfiguration to improve distributed
  generation hosting capacity in active distribution systems,'' \emph{IEEE
  Transactions on Power Systems}, vol.~30, no.~1, pp. 346--356, 2015.

\bibitem{Baran1989-2}
M.~E. Baran and F.~F. Wu, ``{Network reconfiguration in distribution systems
  for loss reduction and load balancing},'' \emph{IEEE Transactions on Power
  Delivery}, vol.~4, no.~2, 1989.

\bibitem{NIKKHAH2023}
S.~Nikkhah, A.~Rabiee, A.~Soroudi, A.~Allahham, P.~C. Taylor, and D.~Giaouris,
  ``{Distributed flexibility to maintain security margin through decentralised
  TSO–DSO coordination},'' \emph{International Journal of Electrical Power \&
  Energy Systems}, vol. 146, p. 108735, 2023.

\bibitem{SYRRI2016}
A.~L. Syrri and P.~Mancarella, ``Reliability and risk assessment of
  post-contingency demand response in smart distribution networks,''
  \emph{Sustainable Energy, Grids and Networks}, vol.~7, pp. 1--12, 2016.

\bibitem{MARTINEZCESENA2016}
E.~{Martínez Ceseña} and P.~Mancarella, ``Practical recursive algorithms and
  flexible open-source applications for planning of smart distribution networks
  with demand response,'' \emph{Sustainable Energy, Grids and Networks},
  vol.~7, pp. 104--116, 2016.

\bibitem{c2c}
\BIBentryALTinterwordspacing
{Electricity North West}, ``{The Capacity to Customers project}.'' [Online].
  Available:
  \url{https://www.enwl.co.uk/go-net-zero/innovation/key-projects/c2c/}
\BIBentrySTDinterwordspacing

\bibitem{network_UK}
\BIBentryALTinterwordspacing
E.~Martinez-Ceseña and A.~Churkin, ``{Synthetic electricity distribution
  network from UK for flexibility analysis - ATTEST project [data set]},''
  2022. [Online]. Available: \url{https://doi.org/10.25747/RE7B-ES77}
\BIBentrySTDinterwordspacing

\bibitem{Capitanescu2017}
F.~Capitanescu, ``A relax and reduce sequential decomposition rolling horizon
  algorithm to value dynamic network reconfiguration in smart distribution
  grid,'' in \emph{2017 IEEE PES Innovative Smart Grid Technologies Conference
  Europe (ISGT-Europe)}, 2017.

\bibitem{Lopez2021}
L.~Lopez, A.~Gonzalez-Castellanos, D.~Pozo, M.~Roozbehani, and M.~Dahleh,
  ``{QuickFlex: a Fast Algorithm for Flexible Region Construction for the
  TSO-DSO Coordination},'' in \emph{2021 International Conference on Smart
  Energy Systems and Technologies (SEST)}, 2021.

\end{thebibliography}

\end{document}